\documentclass{llncs}
\usepackage{latexsym}
\begin{document}
\newcommand{\eat}[1]{}
\title{Acceptability with general orderings}
\author{Danny De Schreye, Alexander Serebrenik}
\institute{Department of Computer Science, K.U. Leuven\\
Celestijnenlaan 200A, B-3001, Heverlee,
Belgium\\Email: \{Danny.DeSchreye, Alexander.Serebrenik\}@cs.kuleuven.ac.be}
\maketitle

\begin{abstract}
We present a new approach to termination analysis of logic programs. The
essence of the approach is that we make use of general orderings (instead
of level mappings), like it is done in transformational approaches to logic
program termination analysis, but we apply these orderings directly to the
logic program and not to the term-rewrite system obtained through some 
transformation. We define some variants of acceptability, based on 
general orderings, and show how they are equivalent to LD-termination. 
We develop a demand driven, constraint-based approach to verify
these acceptability-variants.

The advantage of the approach over standard acceptability is that in some 
cases, where complex level mappings are needed, fairly simple orderings
may be easily generated. The advantage over transformational approaches is that
it avoids the transformation step all together.

\end{abstract}

\section{Introduction}
It is not uncommon in research to have different research communities
that tackle a same problem from a very different perspective or using totally
different techniques. In some cases, such communities may co-exist
for many years without much integration, cross-fertilisation or even decent
comparison of the relative merits and drawbacks of competing approaches.

In the context of termination analysis of logic programs, two such 
sub-communities are those who develop and apply ``transformational''
approaches and those working on ``direct'' ones.
 A transformational approach first transforms the logic 
program into an ``equivalent'' term-rewrite system (or, in some cases, into an equivalent functional program). Here, equivalence means that, at the very 
least, the termination of the term-rewrite system should imply the termination
of the logic program, for some predefined collection of queries\footnote{The approach of Arts~\cite{Arts:PhD} is exceptional in the sense that 
the termination of the logic program is concluded from a weaker property
of {\em single-redex normalisation\/} of the term-rewrite system.}. Direct approaches do not include such a transformation, but prove the termination directly
on the basis of the logic program. 

Besides the transformation step itself, there is one other technical difference
between these approaches. Direct approaches usually prove termination on the 
basis of a well-founded ordering over the natural numbers. More specifically,
they use a {\it level mapping}, which maps atoms to
natural numbers, and, they verify appropriate decreases of this level mapping 
on the atoms occurring in the clauses. On the other hand, transformational   
approaches make use of more general well-founded orderings over terms, such as 
reduction orderings, or more specifically simplification orderings, or others 
(see~\cite{Dershowitz:RTA}). 

At least for the direct approaches the systematic choice for level mappings
and {\em norms}---functions which map each term
(module variable renaming) to a corresponding natural number---instead of 
general orderings, seems arbitrary and ad hoc. More
generally, the relative merits and drawbacks of these two lines of work
are not well understood. This has been the main motivation for this paper. 
We present an initial study on the use of general well-founded orderings 
as a means of directly proving the termination of logic programs---without 
intermediate transformation. In particular, 
\begin{itemize}
\item
we study whether the theoretical results on acceptability can be reformulated on the
basis of general orderings,
\item
we evaluate to what extent the use of the general orderings (instead
of level mappings) either improves or deteriorates the direct approaches.
\end{itemize}

To illustrate the latter point, consider the following program, that formulates
some of the rules for computing the repeated derivative of a linear function 
in one variable $u$ (see also~\cite{DM79:cacm}) :
\begin{example}
\label{example:rep:der}
\begin{eqnarray*}
&& \mbox{\sl d}(\mbox{\sl der}(u),1).\\
&& \mbox{\sl d}(\mbox{\sl der}(A),0) \leftarrow \mbox{\sl number}(A).\\
&& \mbox{\sl d}(\mbox{\sl der}(X+Y),DX+DY) \leftarrow \mbox{\sl d}(\mbox{\sl der}(X),DX), \mbox{\sl d}(\mbox{\sl der}(Y),DY).\\
&& \mbox{\sl d}(\mbox{\sl der}(X*Y),X*DY+Y*DX) \leftarrow \mbox{\sl d}(\mbox{\sl der}(X),DX), \mbox{\sl d}(\mbox{\sl der}(Y),DY).\\
&& \mbox{\sl d}(\mbox{\sl der}(\mbox{\sl der}(X)),DDX)\leftarrow \mbox{\sl d}(\mbox{\sl der}(X),DX), \mbox{\sl d}(\mbox{\sl der}(DX),DDX).
\end{eqnarray*}

We are interested in proving LD-termination, i.e.,
finiteness of the SLD-tree constructed using the left-to-right selection
rule of Prolog, of the program above together with the queries of the form
$\mbox{\sl d}(t, v)$ , where $t$ is a term, expressing a derivative
of a linear function in one variable $u$, such as 
$\mbox{\sl der}(\mbox{\sl der}(u*u*u + 3*u*u + 3*u + 1))$,
and $v$ is a fresh variable, that will be unified with the result of
the computation. 

Doing this on the basis of a level-mapping is hard. 
For this example, a level-mapping that decreases between two sequential
calls of $d$ is a non-linear function. 
In particular, a level mapping $\mid\cdot\mid$,
and a norm $\|\cdot\|$, such that:  
$\mid\!\!\mbox{\sl d}(X,Y)\!\!\mid\; = \|X\|,$ 
$\mid\!\!\mbox{\sl number}(X)\!\!\mid = 0,$ 
$\| \mbox{\sl der}(X) \| = 2^{\|X\|}, $ 
$\| X + Y \| = \mbox{\sl max}(\|X\|, \|Y\|) + 1,$ 
$\| X * Y \| = \mbox{\sl max}(\|X\|, \|Y\|) + 1,$ 
$\| u \| = 2,$ 
$\| n \| = 2,\;\;\mbox{\rm if $n$ is a number}$,%
would be needed. No automatic system for
proving termination on the basis of level mappings is able to generate such
mappings. Moreover, we believe, that it would be very difficult to extend 
existing systems to support generation of appropriate non-linear mappings.
$\hfill\Box$\end{example}

Although we have not yet presented our general-well-founded ordering 
approach, it should be intuitively clear, that we can capture the decrease in 
ordering between the $\mbox{\sl der}(X)$ and $DX$  by using an 
ordering on terms that gives the highest ``priority'' to the functor 
{\sl der}. 

On the other hand, using level mappings and norms allows sometimes to explore
more precise information on atoms and terms, that cannot be expressed 
by general orderings, such as arithmetical relations between terms.
This information can sometimes be crucial in proving termination as 
the following program from~\cite{DeSchreye:Decorte:NeverEndingStory,Decorte:DeSchreye:Vandecasteele} demonstrates.
\begin{example}
\label{example:conf}
\begin{eqnarray*}
&& \mbox{\sl conf}(X)\leftarrow \mbox{\sl delete$_2$}(X,Z), \mbox{\sl delete}(U,Y,Z), \mbox{\sl conf}(Y).\\
&& \mbox{\sl delete$_2$}(X,Y)\leftarrow \mbox{\sl delete}(U,X,Z), \mbox{\sl delete}(V,Z,Y).\\
&& \mbox{\sl delete}(X,[X|T],T).\\
&& \mbox{\sl delete}(X,[H|T],[H|T1])\leftarrow \mbox{\sl delete}(X,T,T1). 
\end{eqnarray*}

Note that by reasoning in terms of sizes of terms, we can infer that the size 
decreases by 2 after the call to $\mbox{\sl delete}_2$ predicate in the
first clause and then increases by 1 in the subsequent call to the 
{\sl delete} predicate. In total, sizes allow us to conclude a decrease. 
Reasoning in terms 
of ordering relations only, however, does not allow to conclude the overall 
decrease from the facts that the third argument of {\sl delete} predicate
is smaller (with respect to some $>$) than the second one and  that the 
first argument of $\mbox{\sl delete}_2$ predicate
is greater (with respect to $>$) than the second one.
$\hfill\Box$\end{example}

As can be expected, theoretically both approaches are essentially equivalent.
We will introduce a variant of the notion of acceptability, based on general
orderings, which is again equivalent to termination in a similar way as 
in the level mapping based approach. On the more practical level, as illustrated  in the two examples above, neither of the approaches is strictly better: the
general orderings provide a larger set of orderings to select from (in 
particular, note that orderings based on level mappings and norms are general 
orderings), the level mapping approach provides arithmetic, on top of mere 
ordering. 

In the remainder of this paper, we will start off from a variant of the notion of 
{\it acceptability with respect to a set}, as introduced in \cite{DeSchreye:Verschaetse:Bruynooghe},
obtained by replacing level mappings by orderings. We show how 
this variant of acceptability
remains equivalent to termination under the left-to-right selection rule, for certain goals. 
Then, we illustrate how this result can be used to prove termination with some examples. We also provide a variant of the {\em acceptability} condition, as 
introduced in~\cite{Apt:Pedreschi:Studies}, and discuss advantages and 
disadvantages of each approach. 
Next, we discuss automation of the approach. We elaborate on a demand-driven method to set-up 
and verify sufficient preconditions for termination. In this method, the aim is to 
derive---in, as much as possible, a constructive way---a well-founded ordering over the set 
of all atoms and terms of the language underlying the program, that satisfies the termination 
condition.

\section{Preliminaries}

\subsection{Logic Programs}

We follow the standard notation for terms and atoms. A {\em query} is a 
finite sequence of atoms. Given an atom $A$, $\mbox{\sl rel}(A)$ denotes
the 
predicate occurring in $A$. $\mbox{\sl Term}_P$ and $\mbox{\sl Atom}_P$ 
denote, respectively, sets of all terms and atoms that can be constructed from
the language underlying $P$. The extended Herbrand Universe $U^E_P$ (the 
extended Herbrand base $B^E_P$) is a quotient set of $\mbox{\sl Term}_P$ 
($\mbox{\sl Atom}_P$) modulo the variant relation. 

We refer to an SLD-tree constructed using the left-to-right selection rule of
Prolog, as an LD-tree. We will say that a goal $G$ {\it LD-terminates} for
a program $P$, if the LD-tree for $(P,G)$ is finite.

The following definition is borrowed from~\cite{Apt:Book}.
\begin{definition}
Let $P$ be a program and $p$, $q$ be predicates occurring in it.
\begin{itemize}
\item We say that {\em $p$ refers to $q$ in $P$\/} if there is a clause in $P$
that uses $p$ in its head and $q$ in its body.
\item We say that {\em $p$ depends on $q$ in $P$\/} and write $p\sqsupseteq q$,
if $(p,q)$ is in the transitive, reflexive closure of the relation {\em refers to}.
\item We say that {\em $p$ and $q$  are mutually recursive\/} and write $p\simeq q$, if $p\sqsupseteq q$ and $q\sqsupseteq p$.
\end{itemize}
\end{definition}

\subsection{Quasi-orderings and orderings}
A {\em quasi-ordering} over a set $S$ is a reflexive and transitive
relation $\geq$ defined on elements of $S$.  We define the associated
equivalence relation $\leq\geq$ as $s \leq\geq t$ if and only if $s\geq t$ and $t\geq s$,
and the associated {\em ordering\/} $>$ 
as $s > t$ if and only if $s\geq t$ but
not $t\geq s$. If neither $s\geq t$, nor $t\geq s$ we write
$s\|_> t$. Sometimes, in order to distinguish between different quasi-orderings
and associated relations we also use $\succeq$, $\succ$, $\preceq\succeq$ and 
$\|_\succ$.

An ordered set $S$ is said to be {\em well-founded}
if there are no infinite descending sequences $s_1 > s_2 > \ldots$ of elements
of $S$. If the set $S$ is clear from the context we will say that the ordering,
defined on it, is well-founded. We'll also say that a quasi-ordering is 
well-founded if the ordering associated with it, is well-founded.

\begin{definition}
Let $\geq$ be a quasi-ordering on a set $T$. 
A quasi-ordering $\succeq$ defined on a set $S\supseteq T$ is called a 
{\em proper extension of $\geq$\/} if
\begin{itemize}
\item $t_1 \geq t_2$ implies $t_1\succeq t_2$ for all $t_1, t_2\in T$.
\item $t_1 > t_2$ implies $t_1\succ t_2$ for all $t_1, t_2\in T$.
\end{itemize}
\end{definition}

The study of termination of term-rewriting systems caused intensive study of
orderings on terms. A number of useful properties were established.

\begin{definition}
Let $>$ be an ordering on $U^E_P \cup B^E_P$.
\begin{itemize}
\item $>$ is called {\em monotonic} if $s_1 > s_2$ implies $f({\bar t_1},s_1,{\bar t_2}) > f({\bar t_1},s_2,{\bar t_2})$ and $p({\bar t_1},s_1,{\bar t_2}) > p({\bar t_1},s_2,{\bar t_2})$ 
for any terms $s_1$ and $s_2$, sequences of terms ${\bar t_1}$ and ${\bar t_2}$, 
function symbol $f$ and predicate $p$.
\item $>$ is said to have the {\em subterm property} if $f({\bar t_1},s,{\bar t_2}) > s$ holds
for any term $f({\bar t_1},s,{\bar t_2})$. 
\end{itemize}
\end{definition}

We extend the definition above to quasi-orderings.
\begin{definition}
Let $\geq$ be a quasi-ordering on terms.
\begin{itemize}
\item $\geq$ is called {\em monotonic} if 
\begin{itemize}
\item 
$s_1 \geq s_2$ implies $f({\bar t_1},s_1,{\bar t_2}) \geq f({\bar t_1},s_2,{\bar t_2})$ and 
$p({\bar t_1},s_1,{\bar t_2}) \geq p({\bar t_1},s_2,{\bar t_2})$ 
for any terms $s_1$ and $s_2$, sequences of terms ${\bar t_1}$ and ${\bar t_2}$, 
function symbol $f$ and predicate $p$ and
\item the associated ordering is monotonic.
\end{itemize}
\item $\geq$ is said to have the {\em subterm property} if the associated 
ordering has the subterm property.
\end{itemize}
\end{definition}

The following are examples of orderings: $>$ on the set of numbers,
lexicographic ordering on the set of strings (this is the way the entries are 
ordered in dictionaries), multiset ordering and recursive path 
ordering~\cite{Dershowitz:RTA}. The following are examples of quasi-orderings:
$\geq$ on the set of numbers, $\supseteq$ on the power set of some set.

For our purposes monotonicity and subterm properties are too restrictive.
Thus, we assign to each predicate or functor a subset of argument positions,
such that for the argument positions in this subset the specified properties
hold. We will say that a predicate $p$ (a functor $f$) is monotone (has a 
subterm property) on a specified subset of argument positions. The formal
study of these weaker notions may be found in~\cite{Serebrenik:DeSchreye:cw291}.

\begin{example}
Let $f$ be a functor of arity two, and $a$, $b$ two terms, such that 
$a > b$. Let $f$ be monotone in the first argument position. Then,
$f(a,c) > f(b,c)$ holds for any term $c$, but there might be some term
$c$, such that $f(c,a) \not > f(c,b)$.
\end{example}

\section{Order-acceptability with respect to a set}
In this section we present and discuss some of the theory we developed
to extend acceptability to general orderings. In the literature, there are 
different variants of acceptability. The most well-known of these is the 
acceptability as introduced by Apt and Pedreschi~\cite{Apt:Pedreschi:Studies}.
This version is defined and verified on the level of ground instances of
clauses, but draws its practical power mostly from the fact that termination is
proved for {\em any bounded\/} goal. Here, boundedness is a notion related to
the selected level mapping and requires that the set $\{|G\theta|\;\mid\;\theta
\;\mbox{\rm is a grounding substitution for goal}\;G\}$ is bounded in the 
natural numbers, where $|\cdot|: B_P\rightarrow {\cal N}$ denotes the level 
mapping.

Another notion of acceptability is the ``acceptability with respect to a set 
of goals'', introduced in~\cite{DeSchreye:Verschaetse:Bruynooghe}. This notion allows to prove termination with respect to any set of
goals of interest. However, it relies on procedural concepts, such as calls
and computed answer substitution. It was designed to be verified through
global analysis, for instance through abstract interpretation.

A variant of acceptability with respect to a set that avoids the drawbacks of using 
procedural notions and that can be verified on a local level was designed
in~\cite{Decorte:DeSchreye:Vandecasteele}. This variant required that the 
goals of interest are {\em rigid\/} under the given level mapping. Here, 
rigidity means that $|G\theta| = |G|$, for any substitution $\theta$, where
$|\cdot|: B^E_P\rightarrow {\cal N}$ now denotes a generalised level mapping,
defined on the extended Herbrand base.

Comparing the notions of boundedness and rigidity in the context of a level 
mapping based approach, it is clear that boundedness is more general than 
rigidity. If the level mapping of a goal is invariant under substitution, then the level mapping is bounded on the set of instances of the goal, but not 
conversely.

Given the latter observation and given that acceptability of~\cite{Apt:Pedreschi:Studies} is a more generally known and accepted notion, we started our work 
by generalising this variant.

However, it turned out that generalising the concept of boundedness to general 
orderings proved to be very difficult. We postpone the discussion on this 
issue until after we formulated the results, but because of these complications, we only arrived at generalised acceptability conditions that are useful
in the context of well-moded and simply moded programs and goals.

Because of this, we then turned our attention to acceptability
with respect to a set. Here, the generalisation of rigidity was less 
complicated, so that in the end we obtained the strongest results for this
variant of acceptability. Therefore, we first present order-acceptability with
respect to a set of goals. We need the following notion.

\begin{definition}~\cite{Decorte:DeSchreye:98} 
Let $P$ be a definite program and $S$ be a set of atomic queries. 
The {\em call set}, $\mbox{\sl Call}(P,S)$, is the set of all atoms $A$,
such that a variant of $A$ is a selected atom in some derivation for
$P\cup \{\leftarrow Q\}$, for some $Q\in S$ and under the left-to-right 
selection rule.
\end{definition}

To illustrate this definition recall the following 
example~\cite{Apt:Book,Decorte:DeSchreye:Vandecasteele}.

\begin{example}
\label{example:permute}
\begin{eqnarray*}
&& {\mbox {\sl permute}}([],[]).\\
&& {\mbox {\sl permute}}(L,[El|T])\leftarrow {\mbox {\sl delete}}(El,L,L1), {\mbox {\sl permute}}(L1,T).\\
&& {\mbox {\sl delete}}(X,[X|T],T).\\
&& {\mbox {\sl delete}}(X,[H|T],[H|T1])\leftarrow {\mbox {\sl delete}}(X,T,T1).
\end{eqnarray*}
Let $S$ be $\{{\mbox {\sl permute}}(t_1, t_2)|\;t_1\;\mbox{\rm is a nil-terminated list and}\;t_2\;\mbox{\rm is a free variable}\}$. 
Then, $\mbox{\sl Call}(P,S) =$
\[S \cup \{{\mbox{\sl delete}}(t_1,t_2,t_3)|\;t_1, t_3\;\;\mbox{\rm are free variables and}\;t_2\;\mbox{\rm is a nil-terminated list}\}.\] 
Such information about $S$ could for instance be expressed in terms of the 
rigid types of Janssens and Bruynooghe~\cite{Janssens:Bruynooghe} and 
$\mbox{\sl Call}(P,S)$ could be 
computed using the type inference of~\cite{Janssens:Bruynooghe}. 
 $\hfill\Box$\end{example}

The following definition generalises the notion of acceptability with respect to 
a set~\cite{Decorte:DeSchreye:98} in two ways: 1) it generalises it to general
quasi-orderings, 2) it generalises it to mutual recursion, using
 the standard notion of mutual recursion~\cite{Apt:Book}.
\begin{definition}
\label{def:taset}
Let $S$ be a set of atomic queries and $P$ a definite 
program. $P$ is {\em order-acceptable with respect to  $S$} if there exists a 
well-founded quasi-ordering $\geq$, such that
\begin{itemize}
\item for any $A\in\mbox{\sl Call}(P,S)$ 
\item for any clause $A'\leftarrow B_1,\ldots,B_n$ in $P$, such that 
$\mbox{\rm mgu}(A,A') = \theta$ exists,
\item for any atom $B_i$, such that $\mbox{\sl rel}(B_i)\simeq \mbox{\sl rel}(A)$
\item for any computed answer substitution $\sigma$ for 
$\leftarrow (B_1, \ldots, B_{i-1})\theta$:
\[A > B_i\theta\sigma.\]
\end{itemize}
\end{definition}

The following establishes the connection between order-acceptability with 
respect to a set $S$ and  LD-termination for queries in $S$. 

\begin{theorem}
\label{taset:term}
Let $P$ be a program. $P$ is order-acceptable with respect to  a set of atomic
queries $S$ if and only if $P$ is LD-terminating for all queries in $S$.
\end{theorem}
\begin{proof}
For all proofs we refer to~\cite{Serebrenik:DeSchreye:cw291}.
\end{proof}

We postpone applying the Theorem~\ref{taset:term} to
Example~\ref{example:permute} until a more syntactic way of
verifying order-acceptability with respect to a set is developed.

To do this, we extend the 
sufficient condition of~\cite{Decorte:DeSchreye:Vandecasteele}, that imposes
the additional requirement of rigidity of the level mapping on the call set,
to the case of general quasi-orderings.

First we adapt the notion of rigidity to general orderings.
\begin{definition}(see also~\cite{Bossi:Cocco:Fabris})
The term or atom $A\in U^E_P \cup B^E_P$ is called {\em rigid} with respect to 
a quasi-ordering $\geq$ if for any substitution $\theta$, $A \leq\geq A\theta$. 
In this case $\geq$ is said to be {\em rigid on} $A$.
\end{definition}

The notion of the rigidity on a term (an atom) is naturally extended to 
the notion of rigidity on a set of atoms (terms). In particular, we will be 
interested in quasi-orderings that are rigid on 
$\mbox{\sl Call}(P,S)$ for some $P$ and $S$. 

We also need interargument relations based on general orderings.
\eat{
\begin{definition}
Let $P$ be a definite program, $p$ a predicate in $P$ with arity $n$ 
and $\geq$ a
quasi-ordering on $U^E_P$. An {\em interargument relation} is a relation 
$R_p = \{(t_1,\ldots,t_n)\mid \varphi_p(t_1, \ldots, t_n)\}$, where:
\begin{itemize}
\item $\varphi_p(t_1,\ldots,t_n)$ is a formula in a disjunctive normal form
\item each conjunct in $\varphi_p$ is either $s_1 > s_2$, $s_1 \leq\geq s_2$ or $s_1\|_> s_2$, where
$s_i$ are constructed from $t_1,\ldots,t_n$ by applying functors of $P$.
\end{itemize}
$R_p$ is a {\em valid interargument relation for $p/n$ with respect to an ordering $>$}
if and only if for every $p(t_1,\ldots,t_n)\in \mbox{\sl Atom}_P\;\mbox{\rm : if}\;\;
P\models p(t_1,\ldots,t_n)$ then $(t_1,\ldots,t_n)\in R_p$.
\end{definition}
}
\begin{definition}
Let $P$ be a definite program, $p$ a predicate in $P$ with arity $n$. 
An {\em interargument relation} is a relation 
$R_p \subseteq \{p(t_1,\ldots,t_n) \mid\;t_i\in \mbox{\sl Term}_P\}$. 
$R_p$ is a {\em valid interargument relation for $p$}
if and only if for every $p(t_1,\ldots,t_n)\in \mbox{\sl Atom}_P\;\mbox{\rm : if}\;\;
P\models p(t_1,\ldots,t_n)$ then $p(t_1,\ldots,t_n)\in R_p$. 
\end{definition}

Usually, the interargument relation will be defined based on a quasi-ordering
used for proving termination. However, in general, this need not be the case.

\begin{example}
Consider the following program.
\begin{eqnarray*}
&& p(0,[]).\\
&& p(f(X),[X|T])\leftarrow p(X,T).
\end{eqnarray*}

The following interargument relations can be considered for $p$:
$\{p(t_1, t_2)\mid t_2 > t_1 \vee t_1 \leq\geq t_2\}$, valid  if 
$\geq$ is a quasi-ordering imposed by a list-length norm, $\|\cdot\|_l$. 
Recall, that for lists $\|[t_1|t_2]\|_l = 1 + \|t_2\|_l$, while the 
list-length of other terms is considered to be 0. 
On the other hand, $\{p(t_1, t_2)\mid t_1 > t_2 \vee t_1 \leq\geq t_2\}$ is 
valid, if $\geq$ is a quasi-ordering imposed by a term-size norm. 

Using general (non-norm based) quasi-orderings, $\{p(t_1, t_2)\mid t_1 > t_2\}$
is valid, for example, for the recursive path ordering~\cite{Dershowitz:RTA} 
with the following ordering on functors: $f/1 \succ ./2$, where
$./2$ is a function symbol defining lists,  and $0 \succ []$. 
Alternatively,
$\{p(t_1, t_2)\mid t_2 > t_1\}$ is valid, for example, for the recursive path 
ordering with the following ordering on functors: $./2 \succ f/1$ and 
$[] \succ 0$.
$\hfill\Box$\end{example}

Using the notion of rigidity we state a sufficient condition for 
order-acceptability with respect to a set.

\begin{theorem}(rigid order-acceptability with respect to $S$)
\label{rigid:acceptability}
Let $S$ be a set of atomic queries and $P$ be a definite program.
Let $\geq$ be a quasi-ordering on $U^E_P$ and for each predicate $p$ in $P$, let $R_p$ be
a valid interargument relation for $p$. If there exists
a well-founded proper extension $\succeq$ of $\geq$ to $U^E_P \cup B^E_P$, which is rigid on 
$\mbox{\sl Call}(P,S)$ such that
\begin{itemize}
\item for any clause $H\leftarrow B_1,\ldots,B_n \in P$, and
\item for any atom $B_i$ in its body, such that 
$\mbox{\sl rel}(B_i)\simeq \mbox{\sl rel}(H)$,
\item for any substitution $\theta$, such that the arguments of the atoms in
$(B_1,\ldots,B_{i-1})\theta$ all satisfy their associated interargument
relations $R_{\mbox{\sl rel}(B_1)},\ldots, R_{\mbox{\sl rel}(B_{i-1})}$
\end{itemize}
\[H\theta \succ B_i\theta\]
then $P$ is order-acceptable with respect to $S$.
\end{theorem}

The stated condition is sufficient for order-acceptability, but is not
necessary for it. Indeed, consider the following example:
\begin{example}
\begin{eqnarray*}
&& p(X) \leftarrow q(X,Y), p(Y).\\
&& q(a,b).
\end{eqnarray*}
Query $\leftarrow p(X)$ terminates with respect to this program. Thus, 
Theorem~\ref{taset:term} implies the program is order-acceptable with respect to
$\{p(X)\}$. However, the conditions of Theorem~\ref{rigid:acceptability} do not
hold. If $\geq$ is a quasi-ordering that satisfies these conditions,
then $p(a) \leq\geq p(b)$ is implied by rigidity and $p(a) > p(b)$ is implied
by the decrease, contradicting the definition of $>$.
\end{example}

We continue the analysis of Example~\ref{example:permute} and show how 
Theorem~\ref{rigid:acceptability} is used.
\begin{example}
Let $\succeq$ be a well-founded quasi-ordering on $U^E_P\cup B^E_P$, such that:
\begin{itemize}
\item for all terms $t_1, t_{21}$ and $t_{22}$:
${\mbox{\sl permute}}(t_1, t_{21}) \preceq\succeq {\mbox{\sl permute}}(t_1, t_{22})$.
\item for all terms $t_{11}, t_{12}, t_2, t_{31}, t_{32}$:
${\mbox{\sl delete}}(t_{11}, t_{2}, t_{31}) \preceq\succeq {\mbox{\sl delete}}(t_{12}, t_{2}, t_{32})$.
\item for all terms $t_{11}, t_{12}$ and $t_2$:
$[t_{11}| t_2] \preceq\succeq [t_{12}| t_2]$.
\end{itemize}

That is, we impose that the quasi-ordering is invariant on predicate argument 
positions and functor argument positions that may occur with a free variable in
${\mbox {\sl Call}}(P,S)$. Furthermore, we impose that $\succeq$ 
has the subterm
and monotonicity properties at all remaining predicate or functor argument
positions. 

First we investigate the rigidity of $\succeq$ on 
${\mbox {\sl Call}}(P,S)$, namely:
$G\theta \preceq\succeq G$ for any $G\in{\mbox {\sl Call}}(P,S)$ and any $\theta$. Now
any effect that the application of $\theta$ to $G$ may have on $G$ needs to be 
through the occurrence of some variable in $G$. However, because we imposed that
$\succeq$ is invariant on all predicate and functor argument positions that may 
possibly contain a variable in some call, $G\theta \preceq\succeq G$.

Associate with ${\mbox{\sl delete}}$ the interargument relation 
$R_{\mbox{\sl delete}} = \{\mbox{\sl delete}(t_1, t_2, t_3)\mid t_2 \succ t_3\}$.
First, we verify that this interargument relationship is valid. Note, that
an interargument relationship is valid whenever it is a model for its 
predicate.
Thus, to check whether $R_{\mbox{\sl delete}}$ is valid, 
$T_P(R_{\mbox{\sl delete}})\subseteq R_{\mbox{\sl delete}}$ is checked.
For the non-recursive clause of ${\mbox{\sl delete}}$ the inclusion follows
from the subset property of $\succeq$, while for the recursive one, from the 
monotonicity of it.

Then, consider the recursive clauses of the program. 
\begin{itemize}
\item ${\mbox{\sl permute}}$. If ${\mbox{\sl delete}}(El,L,L1)\theta$ 
satisfies $R_{\mbox{\sl delete}}$, then $L\theta \succ L1\theta$.
By the monotonicity, ${\mbox{\sl permute}}(L,T)\theta \succ {\mbox{\sl permute}}(L1,T)\theta$. By the property stated above,
 ${\mbox{\sl permute}}(L,[El|T])\theta \preceq\succeq {\mbox{\sl permute}}(L,T)\theta$.
Thus, the desired decrease 
${\mbox{\sl permute}}(L,[El|T])\theta \succ {\mbox{\sl permute}}(L1,T)\theta$
holds.
\item ${\mbox{\sl delete}}$. 
By the properties of $\succ$ stated above:
${\mbox{\sl delete}}(X,[H|T],[H|T1]) \succ {\mbox{\sl delete}}(X,T,[H|T1])$
and ${\mbox{\sl delete}}(X,T,[H|T1]) \preceq\succeq {\mbox{\sl delete}}(X,T,T1)$.
Thus,
${\mbox{\sl delete}}(X,[H|T],[H|T1]) \succ {\mbox{\sl delete}}(X,T,T1)$. 
\end{itemize}

We have shown that all the conditions of Theorem~\ref{rigid:acceptability}
are satisfied, and thus, $P$ is order-acceptable with respect to $S$. 
By Theorem~\ref{taset:term}, $P$ terminates for all queries in $S$.

Observe, that we do not need to construct the actual ordering, but only
to prove that there is one, that meets all the requirements posed. In this 
specific case, the requirement of subterm and monotonicity on the remaining
argument positions is satisfiable.
$\hfill\Box$\end{example}

\section{The results for acceptability with respect to a model}
In this section we briefly discuss some of the results we obtained in 
generalising the acceptability notion of~\cite{Apt:Pedreschi:Studies,%
Etalle:Bossi:Cocco}. Since these results are weaker than those presented 
in the previous section, we do not elaborate on them in full detail. 
\eat{
In particular, we do not
recall the definitions of well-moded programs and goals, nor those of
simply moded programs and goals, that we use below, 
but instead refer to~\cite{Apt:Book}, respectively~\cite{Apt:Etalle}.
}

For a predicate $p$ with arity $n$, a {\em mode} is an atom 
$p(m_1,\ldots, m_n)$, where
$m_i\in \{\mbox{\sl in},\mbox{\sl out}\}$ for $1\leq i\leq n$. Positions with 
$\mbox{\sl in}$ are called {\em input positions}, 
and positions with $\mbox{\sl out}$ are called {\em output positions} of $p$. 
We assume that a fixed mode is associated with each predicate in a program. To
simplify the notation, an atom written as $p({\bf s},{\bf t})$ means: {\bf s}
is the vector of terms filling the input positions, and {\bf t} is the vector
of terms filling the output positions. Furthermore, by 
$\mbox{\sl Var}({\bf s})$ we denote the set of variables occuring in vector of
terms {\bf s}~\cite{Apt:Book}.

Below, we assume that modes for the program and goal are given.
For any atom $A$ and a mode $m_A$ for $A$, we denote by $A^{\mbox{\sl inp}}$ 
the atom obtained from $A$ by removing all output arguments. 
E.g., let $A = p(f(2), 3, X)$ and $m_A = p(\mbox{\sl in},\mbox{\sl in}, 
\mbox{\sl out})$, then $A^{\mbox{\sl inp}} = p(f(2), 3)$.

\begin{definition}
Let $\geq$ be a quasi-ordering relation on $B^E_P$. We say that $\geq$ 
is {\em output-independent\/} if for any two moded atoms $A$ and $B$: 
$A^{\mbox{\sl inp}} = B^{\mbox{\sl inp}}$ implies $A \leq\geq B$.
\end{definition}

The first class of the programs we consider, are {\sl well-moded\/} programs.

\begin{definition}~\cite{Apt:Book} 
\begin{enumerate}
\item 
A query $p_1({\bf s_1},{\bf t_1}),\ldots,p_n({\bf s_n},{\bf t_n})$ is called
{\em well-moded} if for $i\in [1,n]$
\[\mbox{\sl Var}({\bf s_i})\subseteq \bigcup_{j=1}^{i-1}\mbox{\sl Var}({\bf t_j}).\]
\item A clause $p_0({\bf t_0},{\bf s_{n+1}})\leftarrow p_1({\bf s_1},{\bf t_1}),\ldots,p_n({\bf s_n},{\bf t_n})$ is called {\em well-moded} if for 
$i\in [1,n+1]$
\[\mbox{\sl Var}({\bf s_i})\subseteq \bigcup_{j=0}^{i-1}\mbox{\sl Var}({\bf t_j}).\]
\item A program is called {\em well-moded} if every clause of it is.
\end{enumerate}
\end{definition}

For well-moded programs, order-acceptability in the style 
of~\cite{Apt:Pedreschi:Studies} can now be defined as follows.

\begin{definition}
Let $P$ be a well-moded program, $\geq$ an output-independent well-founded 
quasi-ordering and $I$ a model for $P$. 
The program $P$ is called {\em order-acceptable 
with respect to  $\geq$ and $I$\/} if for all $A\leftarrow B_1,\ldots,B_n$ in $P$ and 
all substitutions $\theta$, such that $(A\theta)^{\mbox{\sl inp}}$ and
$B_1\theta,\ldots,B_{i-1}\theta$ are ground and 
$I\models B_1\theta\wedge\ldots \wedge B_{i-1}\theta$ holds: 
$A\theta > B_i\theta$.
\end{definition}

$P$ is called {\em order-acceptable\/} if it is order-acceptable with
respect to some output-independent well-founded quasi-ordering and some model. 
Note the similarity and the difference with the notion of 
{\em well-acceptability} introduced by Etalle, Bossi and 
Cocco~\cite{Etalle:Bossi:Cocco}---both notions relay on ``ignoring'' the output
positions. However, the approach suggested in~\cite{Etalle:Bossi:Cocco} 
measures
atoms by level-mappings, while our approach is based on general 
orderings. In addition~\cite{Etalle:Bossi:Cocco} requires 
a decrease only between atoms of mutually recursive predicates. Similarly,
one might use the notion of order-acceptability that requires a decrease
only between atoms of mutually recursive predicates. This definition will
be equivalent to the one we used, since for atoms of non-mutually recursive 
predicates the dependency relation, $\sqsupset$,
 can always be used to define an ordering.
Since every level mapping naturally gives rise to 
the ordering on atoms, that is $A_1 \succ A_2$ if 
$\mid A_1\mid\;\;>\;\;\mid A_2\mid$, 
we conclude that {\em every well-acceptable program is order-acceptable}.

The following theorem states that order-acceptability of a well-moded program 
is sufficient for termination of well-moded goals with respect to  this 
program. Etalle, Bossi and Cocco~\cite{Etalle:Bossi:Cocco} call such a 
program {\em well-terminating}.

\begin{theorem}
\label{order-acc:term}
Let $P$ be a well-moded program, that is order-acceptable with respect to an 
output-independent well-founded quasi-ordering $\geq$ and a model $I$. 
Let $G$ be a well-moded goal, then $G$ LD-terminates.
\end{theorem}

Note that if the requirement of well-modedness 
of the program $P$ is dropped then the theorem
no longer holds.
\begin{example}
\label{pa:example}
\begin{eqnarray*}
&& p(a) \leftarrow q(X).\\
&& q(f(X)) \leftarrow q(X).
\end{eqnarray*}
We assume the modes $p(\mbox{\sl in})$ and $q(\mbox{\sl in})$ to be given.
This program is not well-moded with respect to  the given modes, because $p(a)$
calls $q/1$ with a free variable, but it 
satisfies the remaining conditions of order-acceptability with respect to 
the following quasi-ordering $\geq$ on terms
$p(a) > q(t)$ and $q(f(t)) > q(t)$ for any term $t$
and $t \leq\geq s$ only if $t$ and $s$ are syntactically identical,
and the following model $I = \{p(a), q(a), q(f(a)), q(f(f(a))), \ldots\}$.
However, note that the well-moded goal $p(a)$ is non-terminating.
$\hfill\Box$\end{example}

Unfortunately, well-modedness is not sufficient to make the converse to hold.
That is, there is a well-moded program $P$ and a well-moded goal $G$, such that
$G$ is LD-terminating with respect to  $P$, but $P$ is not order-acceptable.  
\begin{example}
Consider the following program
\begin{eqnarray*}
&& p(f(X)) \leftarrow p(g(X)).
\end{eqnarray*}
with the mode $p(\mbox{\sl out})$. This program is well-moded, 
the well-moded goal $p(X)$ 
terminates with respect to  this program, but it is not order-acceptable, 
since the required decrease $p(f(X)) > p(g(X))$ violates 
output-independence of $\geq$.
$\hfill\Box$\end{example}

Intuitively, the problem in the example occured, because some information
has been passed via the output positions, i.e, $P$ is not {\em simply
moded}. 

\begin{definition}~\cite{Apt:Etalle}
\begin{enumerate}
\item 
A query $p_1({\bf s_1},{\bf t_1}),\ldots,p_n({\bf s_n},{\bf t_n})$ is called
{\em simply moded} if ${\bf t_1},\ldots,{\bf t_n}$ is a linear family of 
variables and for $i\in [1,n]$
\[\mbox{\sl Var}({\bf s_i})\cap (\bigcup_{j=i}^{n}\mbox{\sl Var}({\bf t_j})) = \emptyset.\]
\item A clause $p_0({\bf s_0},{\bf t_0})\leftarrow p_1({\bf s_1},{\bf t_1}),\ldots,p_n({\bf s_n},{\bf t_n})$ is called {\em simply moded} if $p_1({\bf s_1},{\bf t_1}),\ldots,p_n({\bf s_n},{\bf t_n})$ is simply moded and
\[\mbox{\sl Var}({\bf s_0})\cap (\bigcup_{j=1}^{n}\mbox{\sl Var}({\bf t_j}))=\emptyset.\]
\item A program is called {\em simply moded\/} if every clause of it is.
\end{enumerate}
\end{definition}

Indeed, if $P$ is simply moded the second direction 
of the theorem holds as well. This was already observed 
in~\cite{Etalle:Bossi:Cocco} in 
the context of well-acceptability and well-termination. The following
is an immediate corollary to Theorem 5.1 in~\cite{Etalle:Bossi:Cocco}. As 
that theorem states for well-moded simply moded programs, well-termination
implies well-acceptability. Therefore, well-terminating programs are 
order-acceptable.

\begin{corollary}
\label{term:order-acc}
Let $P$ be a well-moded simply moded program, LD-terminating for any
well-moded goal. Then there exists a
model $I$ and an output-independent well-founded quasi-ordering $\geq$, 
such that $P$ is order-acceptable with respect to  $I$ and $\geq$.
\end{corollary}

To conclude, we briefly discuss why it is difficult to extend the notions 
of order-acceptability to the non well-moded case, using a notion of  
boundedness, as it was done for standard acceptability~\cite{Apt:Pedreschi:Studies}. 
In acceptability based on level mappings, boundedness ensures that the 
level mapping of a (non-ground)
goal can only increase up to some finite bound when the goal becomes more
instantiated. Observe that every ground goal is trivially bounded. 

The most naive approach to generalisation of boundedness is replacing
comparisons of level mappings with orderings, that is defining an atom
$A$ to be {\em bounded\/} with respect to an ordering $>$, if there exists an
atom $C$ such that for all ground instances $A\theta$ of $A$, $C > A\theta$.
Unfortunately, this definition is too week to impose termination.
\begin{example}
\begin{eqnarray*}
&& q\leftarrow p(X).\\
&& p(f(X))\leftarrow p(X).\\
&& p(a).
\end{eqnarray*}
Goal $p(X)$ is bounded with respect to
the quasi-ordering such that $q > \ldots > p(f(f(a))) > p(f(a)) > p(a)$.
Similarly, the decrease requirement between the head and the subgoals is
satisfied, however the goal does not terminate.
\end{example}

Intuitively, the problem in this example occured due to the fact that
infinitely many different atoms are smaller than the boundary. One 
can try to fix this problem by redefining boundedness as:

An atom $A$ is {\em bounded\/} with respect to an ordering $>$, if there exists
an atom $C$ such that for all ground instances $A\theta$ of $A$: $A\theta < C$,
and $\{B\in B^E_P\mid B < C\}$ is finite. 

Such a definition imposes constraints which are very similar to the ones
imposed by standard boundedness in the context of level mappings. However, one
thing we loose is that it is no longer a generalisation of groundness. 
Consider an atom $p(a)$ and assume that our language contains a functor $f/1$ 
and a constant $b$. Then one particular well-founded ordering is
\[p(a) > \ldots > p(f(f(b))) > p(f(b)) > p(b).\]
So, $p(a)$ is not bounded with respect to this ordering.

Because of such complications, we felt that the rigidity-based results
of the previous section are the preferred generalisations to general orderings.

\section{A methodology for verifying order-acceptability}
In this section we present an approach leading towards 
automatic verification of the 
order-acceptability condition. The basic idea for the 
approach is inspired on the ``constraint based'' termination analysis proposed
in~\cite{Decorte:DeSchreye:Vandecasteele}. We start off from the conditions
imposed by order-acceptability, and systematically reduce these conditions
to more explicit constraints on the objects of our search: the quasi-ordering $\geq$ 
and the interargument relations, $R_p$, or model $I$.

The approach presented below 
has been applied successfully to a number of examples that
appear in the literature on termination, such as different versions of
{\sl permute}~\cite{Arts:Zantema:95,Krishna:Rao,Decorte:DeSchreye:Vandecasteele}, {\sl dis-con}~\cite{DeSchreye:Decorte:NeverEndingStory}, {\sl transitive closure}~\cite{Krishna:Rao}, {\sl add-mult}~\cite{Plumer:Book}, {\sl combine}, {\sl reverse}, {\sl odd-even}, {\sl at\_least\_double} and {\sl normalisation}~\cite{Decorte:DeSchreye:Vandecasteele}, {\sl quicksort} program~\cite{Sterling:Shapiro,Apt:Book}, {\sl derivative}~\cite{DM79:cacm}, {\sl distributive law}~\cite{Dershowitz:Hoot}, {\sl boolean ring}~\cite{Hsiang},  {\sl aiakl}, {\sl bid}~\cite{Maria:Benchmarks}, 
credit evaluation expert system~\cite{Sterling:Shapiro},  
{\sl flatten}~\cite{Arts:PhD}, vanilla meta-interpreter {\sl solve}~\cite{Sterling:Shapiro} together with wide class of interpreted programs.

In the remainder of the paper, we explain the approach using some 
of these examples. 

We start by showing how the analysis of Example~\ref{example:permute}, 
presented before, can be performed systematically. We stress the main steps 
of a methodology.

\begin{example}
$\geq$ should be rigid on ${\mbox {\sl Call}}(P,S)$. To enforce the rigidity,
$\geq$ should ignore all argument positions in atoms in ${\mbox {\sl Call}}(P,S)$
that might be occupied by free variables, i.e., the second argument
position of ${\mbox {\sl permute}}$ and the first and the third argument 
positions of ${\mbox {\sl delete}}$. Moreover, since the first argument
of ${\mbox {\sl permute}}$ and the second argument of ${\mbox {\sl delete}}$
are general nil-terminated lists, the first argument of $./2$ should be 
ignored as well. 

The decreases with respect to $>$ imposed in the 
order-acceptability with respect to a set $S$ are:
\begin{eqnarray*}
&& {\mbox {\sl delete}}(X, [H|T], [H|T1])\theta > {\mbox {\sl delete}}(X, T, T1)\theta \\
&& {\mbox {\sl delete}}(El, L, L_1)\theta\;\;{\mbox {\rm satisfies}}\;\;R_{\mbox {\sl delete}}\;\;{\mbox {\rm implies}}\\
&& \hspace{1.0cm}{\mbox {\sl permute}}(L,[El|T])\theta > {\mbox {\sl permute}}(L_1,T)\theta
\end{eqnarray*}

To express the rigidity constraints,  we simplify 
each of these conditions by replacing the predicate argument 
positions that should be ignored by some arbitrary term---one of $v_1, v_2, \ldots$. 
The following conditions are obtained:
\begin{eqnarray}
&& {\mbox {\sl delete}}(v_1, [H|T]\theta, v_2) > {\mbox {\sl delete}}(v_3, T\theta, v_4) \\
&& {\mbox {\sl delete}}(El, L, L_1)\theta\;\;{\mbox {\rm satisfies}}\;\;R_{\mbox {\sl delete}}\;\;{\mbox {\rm implies}} \nonumber\\
&& \hspace{1.0cm}{\mbox {\sl permute}}(L\theta, v_1) > {\mbox {\sl permute}}(L_1\theta, v_2)
\end{eqnarray}

Observe that this replacement 
only partially deals with the requirements that the 
rigidity conditions expressed above impose: rigidity on functor arguments
(the first argument of $./2$ should be invariant with respect to the ordering) is not 
expressed. We keep track of such constraints implicitly, and only verify them
at a later stage when additional constraints on the ordering are derived.

For each of the conditions (1) and (2), we have two options on how to enforce it:

Option 1): The decrease required in the condition can be achieved by imposing
some property on $\geq$, which is consistent with the constraints that were 
already imposed on $\geq$ before.

In our example, condition (1) is satisfied by imposing the subterm property 
for the second argument of $./2$ and monotonicity on the second argument
of ${\mbox {\sl delete}}$. The second argument of $./2$ does not belong to
a set of functor argument positions that should be ignored.
Then, $[t_1|t_2] > t_2$ holds for any terms $t_1$ and $t_2$, and by the
monotonicity of $>$ in the second argument of ${\mbox {\sl delete}}$ 
(1) holds. 

In general we can select from a bunch of ordering properties, or even 
specific orderings, that were proposed in the literature. 

Option 2): The required decrease is imposed as a constraint on the 
interargument relation(s) $R$ of the preceding atoms.

In the ${\mbox {\sl permute}}$ example, the decrease 
${\mbox {\sl permute}}(L\theta, t) > {\mbox {\sl permute}}(L_1\theta, t)$ 
cannot directly be achieved by 
imposing some constraint on $>$. Thus, we impose that the underlying decrease
$L\theta > L_1\theta$ should hold for the intermediate body atoms
(${\mbox {\sl delete}}(El, L, L_1)\theta$) that satisfy the interargument
relation $R_{\mbox {\sl delete}}$.

Thus, in the example, the constraint is that 
$R_{\mbox {\sl delete}}$ should be such that for all
${\mbox {\sl delete}}(t_1, t_2, t_3)$ that satisfy $R_{\mbox {\sl delete}}$: 
$t_2 > t_3$. As we have observed, the interargument relation is valid if it
forms a model for its predicate. Thus, one way to constructively verify that 
a valid interargument relation $R_{\mbox {\sl delete}}$ exists,
such that the property $t_2 > t_3$
holds for ${\mbox {\sl delete}}(t_1, t_2, t_3)$ atoms is to simply impose
that $M = \{{\mbox {\sl delete}}(t_1, t_2, t_3)\mid t_2 > t_3\}$ itself is
a model for the ${\mbox {\sl delete}}$ clauses in the program. 

So our new constraint on $R_{\mbox {\sl delete}}$ is that it should 
include $M$. Practically we
can enforce this by imposing that $T_P(M)\subseteq M$ should hold. As shown
in~\cite{Serebrenik:DeSchreye:cw291}, this reduces to the constraints 
``$[t_1|t_2] > t_2$'' and ``$t_2 > t_3$ implies $[t|t_2] > [t|t_3]$''. 
These are again fed into our
Option 1) step, imposing a monotonicity property on the second argument of 
$./2$ for $>$
. At this point the proof is complete.
$\hfill\Box$\end{example}

Recall that we do not need to construct actually the ordering, but only to prove
that there is one, that meets all the requirements posed.

\section{Further examples}
Although the simplicity of the {\sl permute} example makes it a good
choice to clarify our approach it does not well motivate the need for 
general orderings instead of level mappings. Indeed, it is well-known that
{\sl permute} can be dealt with using standard acceptability 
or acceptability with respect to 
a set~\cite{DeSchreye:Decorte:NeverEndingStory}.

In this section we provide a number of additional examples. Most of them
({\sl distributive law}, {\sl derivative} and {\sl solve}) illustrate
the added power of moving to general orderings. After these we present
an alternative version of {\sl permute} in order to discuss an extension
of our approach that deals with interargument relations for conjunctions
of (body-) atoms.

Before presenting the examples we recall once more the main steps of our 
approach. First, given
a program $P$ and a set $S$ of goals, {\em compute the set of calls\/}
${\mbox {\sl Call}}(P,S)$. Janssens and Bruynooghe~\cite{Janssens:Bruynooghe}
show how this can be done through abstract interpretation.
Second, {\em enforce the rigidity of $>$ on 
${\mbox {\sl Call}}(P,S)$}, i.e., ignore all predicate or functor
argument positions that might be occupied by free variables in 
${\mbox {\sl Call}}(P,S)$. Given the set of calls, this step can be performed
in a completely automatic way.
Third, repeatedly {\em construct decreases with respect to $>$},
such that the rigid order-acceptability condition will hold and check if those
can be verified by some of the predefined orderings. While performing this
verification step the trade-off between efficiency and power should be 
considered---using more complex orderings may allow correct reasoning on more
examples but might be computationally expensive.

First, we consider the distributive law program.
This example originated from~\cite{Dershowitz:Hoot}.
\begin{example}
\label{example:dist}
\begin{eqnarray*}
&& \mbox{\sl dist}(x, x).\\
&& \mbox{\sl dist}(x*x, x*x).\\
&& \mbox{\sl dist}(X+Y, U+V) \leftarrow \mbox{\sl dist}(X,U), \mbox{\sl dist}(Y, V).\\
&& \mbox{\sl dist}(X * (Y+Z), T) \leftarrow \mbox{\sl dist}(X*Y + X*Z, T).\\
&& \mbox{\sl dist}((X+Y) * Z, T) \leftarrow \mbox{\sl dist}(X*Z + Y*Z, T).\\
\end{eqnarray*}

Similarly to the repeated derivation example in the introduction, 
no linear norm is
sufficient for proving termination. The simplest norm, we succeeded 
to find, providing a termination proof is the following one:
$\|X * Y\| = \|X\| * \|Y\|$, $\|X + Y\| = \|X\| + \|Y\| + 1$,
$\|x\| = 2$ and the level mapping is $\mid\!\!\mbox{\sl dist}(X,Y)\!\!\mid\; =
\|X\|$. This norm cannot be generated automatically by
 termination analysers we are aware of.

In order to prove termination of a set of queries \[\{\mbox{\sl dist}(t_1, t_2)\;\mid\;t_1\;\mbox{\rm is an expression in a variable $x$ and}
\;t_2\;\mbox{\rm is a free variable}\}\] we use the rigid-acceptability 
condition. First the quasi-ordering, $\geq$, 
we are going to define should be rigid
on a set of calls, i.e., it should ignore the second argument position of
$\mbox{\sl dist}$. Thus, in the 
decreases with respect to $>$ to follow we replace the 
second argument of $\mbox{\sl dist}$ with anonymous terms $v_1, v_2, \ldots$.

\begin{eqnarray*}
&& \mbox{\sl dist}((X + Y)\theta, v_1) > \mbox{\sl dist}(X\theta, v_2)\\
&& \mbox{\sl dist}(X, U)\theta\;\mbox{\rm satisfies $R_{\mbox{\sl dist}}$ implies} \\
&& \hspace{1.0cm}\mbox{\sl dist}((X + Y)\theta, v_1) > \mbox{\sl dist}(Y\theta, v_2)\\
&& \mbox{\sl dist}((X * (Y + Z))\theta, v_1) > \mbox{\sl dist}((X * Y + X * Z)\theta, v_2)\\
&& \mbox{\sl dist}(((X + Y) * Z)\theta, v_1) > \mbox{\sl dist}((X * Z + Y * Z)\theta, v_2)\\
\end{eqnarray*}

The first two decreases are satisfied by any ordering having a subterm property 
for both arguments of $+/2$ and being monotonic with respect to the first argument 
position of $\mbox{\sl dist}$. However, in order to satisfy the later two
we need to use the recursive path ordering (rpo)~\cite{Dershowitz:RTA},
with $*$ preceding $+$ with respect to an ordering on functors. If this ordering is
used, the following holds for any $t_1, t_2$ and $t_3$:
\begin{eqnarray*}
&& t_2 + t_3 > t_2 \\
&& t_1 * (t_2 + t_3) > t_1 * t_2 \\
&& t_2 + t_3 > t_3 \\
&& t_1 * (t_2 + t_3) > t_1 * t_3 \\
&& t_1 * (t_2 + t_3) > t_1 * t_2 + t_1 * t_3\;\;\mbox{\rm (using the properties of rpo)}
\end{eqnarray*}
This proves the third decrease with respect to $>$. The fourth one is proved analogously.
$\hfill\Box$\end{example}

Now we can return to the motivating Example~\ref{example:rep:der},
on computing higher derivatives of polynomial functions in one variable.
\begin{example}
\begin{eqnarray*}
&& \mbox{\sl d}(\mbox{\sl der}(u),1).\\
&& \mbox{\sl d}(\mbox{\sl der}(A),0) \leftarrow \mbox{\sl number}(A).\\
&& \mbox{\sl d}(\mbox{\sl der}(X+Y),DX+DY) \leftarrow \mbox{\sl d}(\mbox{\sl der}(X),DX), \mbox{\sl d}(\mbox{\sl der}(Y),DY).\\
&& \mbox{\sl d}(\mbox{\sl der}(X*Y),X*DY+Y*DX) \leftarrow \mbox{\sl d}(\mbox{\sl der}(X),DX), \mbox{\sl d}(\mbox{\sl der}(Y),DY).\\
&& \mbox{\sl d}(\mbox{\sl der}(\mbox{\sl der}(X)),DDX)\leftarrow \mbox{\sl d}(\mbox{\sl der}(X),DX), \mbox{\sl d}(\mbox{\sl der}(DX),DDX).
\end{eqnarray*}

We are interested in proving termination of the queries that belong to the set
$S = \{\mbox{\sl d}(t_1, t_2)\;\mid\;t_1\;\mbox{\rm is a repeated derivative of a function in a variable $u$ and}$ $t_2\;\mbox{\rm is a free variable}\}$. 
So $S$ consists of atoms of the form $\mbox{\sl d}(\mbox{\sl der}(u), X)$
or $\mbox{\sl d}(\mbox{\sl der}(u * u + u), Y)$ or 
$\mbox{\sl d}(\mbox{\sl der}(\mbox{\sl der}(u + u)), Z)$, etc.
Observe, that
${\mbox {\sl Call}}(P,S)$ coincides with $S$.

We start by analysing the requirements that imposes the rigidity of $\geq$ on
${\mbox {\sl Call}}(P,S)$. First, the second argument position of $d$ 
should be ignored, since it might be occupied by a free variable. Second, 
the first argument position of $d$ is occupied by a ground term. Thus, rigidity
does not pose any restrictions on functors argument positions.

Then, we construct the decreases with respect to $>$ that follow from the 
rigid order-acceptability.
The arguments that should be ignored are replaced by terms $v_1, v_2, \ldots$. 
\begin{eqnarray}
&& \mbox{\sl d}(\mbox{\sl der}(X+Y)\theta , v_1) > \mbox{\sl d}(\mbox{\sl der}(X)\theta, v_2)\\
&& \mbox{\sl d}(\mbox{\sl der}(X),DX)\theta\;\;\mbox{\rm satisfies}\;\;R_{\mbox{\sl d}}\;\;\mbox{\rm implies} \nonumber \\
&& \hspace{1.0cm}\mbox{\sl d}(\mbox{\sl der}(X+Y)\theta, v_1) > \mbox{\sl d}(\mbox{\sl der}(Y)\theta, v_2)\\
&& \mbox{\sl d}(\mbox{\sl der}(X*Y)\theta, v_1) > \mbox{\sl d}(\mbox{\sl der}(X)\theta, v_2)\\
&& \mbox{\sl d}(\mbox{\sl der}(X),DX)\theta\;\;\mbox{\rm satisfies}\;\;R_{\mbox{\sl d}}\;\;\mbox{\rm implies}\nonumber \\
&& \hspace{1.0cm}\mbox{\sl d}(\mbox{\sl der}(X*Y)\theta, v_1) > \mbox{\sl d}(\mbox{\sl der}(Y)\theta, v_2)\\
&& \mbox{\sl d}(\mbox{\sl der}(\mbox{\sl der}(X))\theta, v_1) > \mbox{\sl d}(\mbox{\sl der}(X)\theta, v_2)\\
&& \mbox{\sl d}(\mbox{\sl der}(X),DX)\theta\;\;\mbox{\rm satisfies}\;\;R_{\mbox{\sl d}}\;\;\mbox{\rm implies}\nonumber \\
&& \hspace{1.0cm}\mbox{\sl d}(\mbox{\sl der}(\mbox{\sl der}(X))\theta, v_1) > \mbox{\sl d}(\mbox{\sl der}(DX)\theta, v_2)
\end{eqnarray} 

Conditions (3)-(7) impose monotonicity and subset properties to hold on 
the first argument of $\mbox{\sl d}$.
In order to satisfy condition (8), it is sufficient to prove that 
for any $(t_1, t_2)\in R_{\mbox{\sl d}}$ holds that $t_1 > t_2$.
That is if $M = \{\mbox{\sl d}(t_1, t_2)\mid t_1 > t_2\}$ then 
$T_P(M)\subseteq M$. This may be reduced to the following conditions:
\[
\begin{array}{lc}
\mbox{\sl der}(t) > 1 & (9)\\
t_1\in R_{\mbox{\sl number}}\;\;\mbox{\rm implies}\;\;\mbox{\sl der}(t_1) > 0 & (10)\\
\mbox{\sl der}(t_1) > t_2\;\;\&\;\;\mbox{\sl der}(t_3) > t_4\;\;\mbox{\rm implies}\;\;\mbox{\sl der}(t_1 + t_3) > t_2 + t_4 & (11)\\
\mbox{\sl der}(t_1) > t_2\;\&\;\;\mbox{\sl der}(t_3) > t_4\;\;\mbox{\rm implies}\;\;\mbox{\sl der}(t_1 * t_3) > t_1 * t_4 + t_2 * t_3 & (12)\\
\mbox{\sl der}(t_1) > t_2\;\;\&\;\;\mbox{\sl der}(t_2) > t_3\;\;\mbox{\rm implies}\;\;\mbox{\sl der}(\mbox{\sl der}(t_1)) > t_3 & (13)
\end{array}
\]
Condition (13) follows from monotonicity and transitivity of $>$. However,
(10)-(12) are not satisfied by general properties of $>$ and we need to
specify the ordering. The ordering that meets these conditions is the recursive 
path ordering~\cite{Dershowitz:RTA} with {\sl der} having the highest priority.
$\hfill\Box$\end{example}

As a next example we demonstrate that the suggested technique is 
useful for proving termination of meta-interpreters as well.
\begin{example}
\label{example:solve}
\begin{eqnarray*}
&& \mbox{\sl solve}(true).\\
&& \mbox{\sl solve}((A,B)) \leftarrow  \mbox{\sl solve}(A), \mbox{\sl solve}(B).\\
&& \mbox{\sl solve}(A) \leftarrow  \mbox{\sl clause}(A,B), \mbox{\sl solve}(B).
\end{eqnarray*}

Even though the termination of an interpreted program might be easily
proved with level-mappings, the termination proof of the meta-interpreter
with respect to it cannot be immediately constructed based on the termination 
proof of the interpreted program.

Indeed, let {\sl P} be the interpreted program:
\begin{eqnarray*}
&& \mbox{\sl p}([X,Y|T]) \leftarrow  \mbox{\sl p}([Y|T]),\mbox{\sl p}(T). 
\end{eqnarray*}

Termination of the set of queries $\{\mbox{\sl p}(t)\mid\;
t\;\mbox{\rm is a list of a finite length}\}$ can be easily proved, 
for example by a using level mapping $\mid\!\!\mbox{\sl p}(X)\!\!\mid\; = \|X\|_l$ and
the list-length norm $\|\cdot\|_l$. However, when this program is considered
together with this meta-interpreter these level-mapping and norm cannot
be extended in a way allowing to prove termination, even though there exist
a linear level-mapping and a linear norm that provide a termination proof.
In the case of this example, the following linear level mapping is sufficient
for proving termination: 
\eat{
$\|(A, B)\| = 1+\|A\|+\|B\|$, $\| p(X)\| = 1+\|X\|$, $\|[H|T]\| =
1+3\|T\|$.
}
\begin{eqnarray*}
&& \mid\!\!\mbox{\sl solve}(A)\!\!\mid\; = \| A \| \\
&& \|(A, B)\| = 1+\|A\|+\|B\|\\
&& \| p(X)\| = 1+\|X\| \\
&& \|[H|T]\| = 1+3\|T\|
\end{eqnarray*}

The constraint-based approach of~\cite{Decorte:DeSchreye:Vandecasteele}
is able to derive this level mapping. However, it cannot reuse any information
from a termination proof of the interpreted program to do so, and the 
constraints set up for such examples are fairly complex ($n$ body atoms are
interpreted as a $,/2$-term of depth $n$ and reasoning on them requires
products of (at least) $n$ parameters). Most other approaches based on
level mappings work on basis of fixed norms, like list-length and term-size,
and therefore fail to prove termination of the example.

Applying general orderings allows to define a new ordering
for the meta-interpreter together with the interpreted program based
on the ordering obtained for the interpreted program itself. More formally,
given a quasi-ordering $\geq$, defined for the interpreted program above,
define a quasi-ordering $\succeq$ on terms and atoms of the meta-interpreter,
as follows (similarly to rpo~\cite{Dershowitz:RTA}):
\begin{itemize}
\item $t \preceq\succeq s$ if one of the following holds:
\begin{itemize}
\item $t \leq\geq s$
\item $t = (t_1, t_2), s = (s_1, s_2)$ and $t_1 \preceq\succeq s_1$, $t_2 \preceq\succeq s_2$
\item $t = \mbox{\sl solve}(t_1), s =\mbox{\sl solve}(s_1)$ and $t_1 \preceq\succeq s_1$
\end{itemize}
\item $t\succ s$ if one of the following holds:
\begin{itemize}
\item $t > s$
\item $t = f(\ldots), s = (s_1, s_2)$, $f$ differs from 
$,/2,\mbox{\sl solve}/1$, $t\succ s_1$ and $t\succ s_2$
\item $t = (t_1, t_2)$ and either $t_1\succeq s$ or $t_2\succeq s$.
\item $t = \mbox{\sl solve}(t_1), s =\mbox{\sl solve}(s_1)$ and $t_1 \succ s_1$.
\item $t = \mbox{\sl solve}(t_1), s =\mbox{\sl clause}(s_1, s_2)$
\end{itemize}
\end{itemize}

In our case $\geq$ is a list-length norm based ordering, and $\succeq$ is defined
as specified. Then, $\mbox{\sl p}([X,Y|T]) \succ (\mbox{\sl p}([Y|T]),\mbox{\sl p}(T))$. This provides the $\succ$-decrease for the second recursive 
clause of the meta-interpreter required in the rigid
order-acceptability condition.
Similarly, the decrease for the first recursive clause is provided by the 
subterm property that $\succ$ is defined to have, and thus, proving 
termination.

By reasoning in a similar way, termination can be proved for 
the meta-interpreter and wide class of interpreted programs: from the small 
examples, such as {\sl append} and {\sl delete} and up to 
bigger ones, like {\sl aiakl}, {\sl bid}~\cite{Maria:Benchmarks}, 
credit evaluation expert system~\cite{Sterling:Shapiro}, or even
the {\em distributive law\/} program, presented in Example~\ref{example:dist}. 
$\hfill\Box$\end{example}
\eat{
\begin{example}
\label{example:solve:acc}
The following meta-interpreter appeared in~\cite{Bruynooghe:DeSchreye:Martens}.
Unlike the meta-interpreter, presented in Example~\ref{example:solve}
this meta-interpreter accumulates the goal atoms it will have to resolve.
Thus, potentially, the argument of {\sl solve} may have any number of
subgoals.

\begin{eqnarray*}
&& \mbox{\sl solve}(\mbox{\sl true}).\\
&& \mbox{\sl solve}((G,Gs)) \leftarrow  \mbox{\sl clause}(G,B),\\
&& \hspace{1.0cm} \mbox{\sl concat}(B, Gs, Gs1),\\
&& \hspace{1.0cm} \mbox{\sl solve}(Gs1).
\end{eqnarray*}

augmented with the following {\sl concat} program (we assume that
the goals are always ended with the \mbox{\sl true} atoms).
\begin{eqnarray*}
&& \mbox{\sl concat}(\mbox{\sl true}, Gs, Gs).\\
&& \mbox{\sl concat}((G, Gs1), Gs2, (G, Gs3))\leftarrow \mbox{\sl concat}(Gs1, Gs2, Gs3).
\end{eqnarray*}

Let {\sl P} be the interpreted program:
\begin{eqnarray*}
&& \mbox{\sl p}([X,Y|T]) \leftarrow  \mbox{\sl p}([Y|T]),\mbox{\sl p}(T). 
\end{eqnarray*}

Similarly, to the previous example the natural level mapping for the
interpreted program
$\mid\!p(X)\!\mid\;=\|X\|_l$ and a list-length norm $\|\cdot\|_l$ cannot
be extended to provide a termination proof for the meta-interpreter.

The quasi-ordering $\succeq$ 
that allows us to prove termination of this example is also based
on the quasi-ordering $>$ used for proving termination of the interpreted
programs as follows. First, we define a function $\tau$ that maps
a term $t$ with a primitive functor $,/2$, to a multiset of it 
non $,/2$-subterms. For example, $\tau((a,(b,(c, f(d))))) =\{\!\!\{a,b,c,f(d)\}\!\!\}$.
Then, we extend an ordering $>$ to $\gg$ on multisets and define
$t\succ s$ if $\tau(t)\gg \tau(s)$. Then, given an interargument
relation $\tau((t_1, t_2)) = \tau(t_3)$ for 
$\mbox{\sl concat}(t_1, t_2, t_3)$, termination of the example
follows from the order-acceptability condition.

The same technique can be applied for proving termination 
of meta-interpreter together with a wide class of interpreted programs
mentioned above.
$\hfill\Box$\end{example}
}
The previous examples do not illustrate our approach in full generality. In
general, we may have clauses of the type
\[p(t_1,\ldots,t_n)\leftarrow B_1,B_2,\ldots,B_{i-1},q(s_1,\ldots,s_m),B_{i+1},\ldots,B_k.\]
where multiple intermediate body-atoms, $B_1,B_2,\ldots,B_{i-1}$ precede
the (mutually) recursive body-atom $q(s_1,\ldots,s_m)$. In such cases
the decrease with respect to $>$ between $p(t_1,\ldots,t_n)\theta$ and $q(s_1,\ldots,s_m)\theta$
required by the (rigid) order-acceptability imposes a constraint on
$R_{\mbox{\rm rel}(B_1)}, R_{\mbox{\rm rel}(B_2)},\ldots$ and
$R_{\mbox{\rm rel}(B_{i-1})}$. However, our previous technique of
using $T_P(M)\subseteq M$ to translate the required decrease to
$R_{\mbox{\rm rel}(B_1)}, R_{\mbox{\rm rel}(B_2)},\ldots,
R_{\mbox{\rm rel}(B_{i-1})}$ is not easily generalised. This is because
several of the atoms $B_1,B_2,\ldots,B_{i-1}$ together may be responsible 
for the decrease and the $T_P(M)\subseteq M$ technique is not readily 
generalised to deal with multiple predicates.

One way to deal with this is based on early works on termination analysis
(\cite{Ullman:van:Gelder,Plumer:Book}). Assume that the underlying decrease
imposed by 
\begin{eqnarray*}
&&B_1\theta,B_2\theta,\ldots,B_{i-1}\theta\;\;\mbox{\rm satisfy}\;\;R_{\mbox{\rm rel}(B_1)}, R_{\mbox{\rm rel}(B_2)},\ldots,
R_{\mbox{\rm rel}(B_{i-1})}\;\;\mbox{\rm implies} \nonumber \\
&& \hspace{1.0cm}p(t_1,\ldots,t_n)\theta > q(s_1,\ldots,s_m)\theta
\end{eqnarray*}
is of the form
$u\theta > v\theta$, where $u$ and $v$ are subterms of $p(t_1,\ldots,t_n)$,
respectively $q(s_1,\ldots,s_m)$. We then search for a sequence of terms
$u, u_1, u_2, \ldots, u_j, v$, such that for each pair of terms, $u$ 
and $u_1$, $u_1$ and $u_2$, $\ldots$, $u_j$ and $v$, there is a 
corresponding atom in the sequence $B_1, B_2,\ldots, B_{i-1}$ that contains
both of them.

Assume (without real loss of generality) that $u$ and $u_1$ occur in $B_1$,
$u_1$ and $u_2$ occur in $B_2$, $\ldots$, $u_j$ and $v$ occur in $B_{i-1}$.
We then select one of these pairs of terms, say $u_{i_1}$ and $u_{i_2}$ 
in atom $B_{i_3}$, and impose the relations:
\[
\begin{array}{llll}
u_{i_1} < u_{i_2} &\mbox{\rm on}& R_{\mbox{\rm rel}(B_{i_3})}\mbox{\rm ,} &
\mbox{\rm and}\\
u_{i_1} \leq u_{i_2} &\mbox{\rm on}& R_{\mbox{\rm rel}(B_{i_3})} &
\mbox{\rm for all other pairs of terms and corresponding atoms.}
\end{array}
\]

Now we can again use the $T_P(M)\subseteq M$ technique to translate such 
constraints into interargument relations.

Note that this approach involves a search problem: if we fail to verify 
the proposed inequality constraints, we need to backtrack over the choice
of:
\begin{itemize}
\item the pair $u_{i_1}$ and $u_{i_2}$ in $B_{i_3}$ with a strict inequality, or
\item the sequence of terms $u, u_1, u_2, \ldots, u_j, v$ in $B_1, B_2,\ldots, B_{i-1}$.
\end{itemize}

A completely different method for dealing with multiple intermediate body-atoms
is based on the use of unfold/fold steps to group atoms. We illustrate this
second method with an example.
\begin{example}
\label{example:2intatoms}
The following is the version of the {\sl permute} program that appeared 
in~\cite{Krishna:Rao}.
\[
\begin{array}{ll}
\mbox{\sl perm}([],[]). & \mbox{\sl ap}_1([],L,L).\\
\mbox{\sl perm}(L,[H|T])\leftarrow & \mbox{\sl ap}_1([H|L1],L2,[H|L3])\leftarrow\\
\hspace{0.5cm} \mbox{\sl ap}_2(V,[H|U],L), &\hspace{0.5cm} \mbox{\sl ap}_1(L1,L2,L3).\\
\hspace{0.5cm} \mbox{\sl ap}_1(V,U,W),  & \mbox{\sl ap}_2([],L,L).\\
\hspace{0.5cm}\mbox{\sl perm}(W,T). & \mbox{\sl ap}_2([H|L1],L2,[H|L3])\leftarrow\\
& \hspace{0.5cm}\mbox{\sl ap}_2(L1,L2,L3).
\end{array}
\]

This example is chosen to illustrate applications of Theorem~\ref{order-acc:term} (the 
well-moded case). We would like to prove termination of the goals 
$\mbox{\sl perm}(t_1, t_2)$, where $t_1$ is a ground list and $t_2$ a free 
variable.

Assume the modes 
$\mbox{\sl perm}(\mbox{\sl in},\mbox{\sl out}), \mbox{\sl ap}_1(\mbox{\sl in},\mbox{\sl in},\mbox{\sl out}), \mbox{\sl ap}_2(\mbox{\sl out},\mbox{\sl out},\mbox{\sl in})$.
The order-acceptability imposes, among the others, the following decrease
with respect to $>$:
$I\models \mbox{\sl ap}_2(V,[H|U],L)\theta\wedge\mbox{\sl ap}_1(V,U,W)\theta$ 
implies $\mbox{\sl perm}(L)\theta > \mbox{\sl perm}(W)\theta$. Note that the 
underlying decrease $L\theta > W\theta$ cannot be achieved by reasoning on 
$\mbox{\sl ap}_1/3$ or $\mbox{\sl ap}_2/3$ alone. 

An alternative solution to the one described before is to use the unfold/fold 
technique to provide a definition for the conjunction of the two intermediate
body-atoms. To do this, we start of from a generalised clause, containing
the conjunction of atoms both in its head and in its body. In our example
we get
\[
\mbox{\sl ap}_2(V,[H|U],L),\mbox{\sl ap}_1(V,U,W)\leftarrow 
\mbox{\sl ap}_2(V,[H|U],L),\mbox{\sl ap}_1(V,U,W)
.\]

Next, we unfold both body-atoms , using all applicable clauses, for one
resolution step. This gives rise to a generalised program $P'$, defining
the conjunction of intermediate body-atoms:
\eat{
\begin{eqnarray*}
&&\mbox{\sl ap}_2([],[t_1|t_2],[t_1|t_2]), \mbox{\sl ap}_1([],t_2,t_2).\\
&&\mbox{\sl ap}_2([t_6|t_1],[t_5|t_2],[t_6|t_3]),\mbox{\sl ap}_1([t_6|t_1],t_2,[t_6|t_4])\leftarrow \\
&& \hspace{1.0cm}\mbox{\sl ap}_2(t_1,[t_5|t_2],t_3), \mbox{\sl ap}_1(t_1,t_2,t_4).
\end{eqnarray*}
}
\begin{eqnarray*}
&&\mbox{\sl ap}_2([],[H|T],[H|T]), \mbox{\sl ap}_1([],T,T).\\
&&\mbox{\sl ap}_2([H1|T1],[H2|T2],[H1|T3]),\mbox{\sl ap}_1([H1|T1],T2,[H1|T4])\leftarrow \\
&& \hspace{1.0cm}\mbox{\sl ap}_2(T1,[H2|T2],T3), \mbox{\sl ap}_1(T1,T2,T4).
\end{eqnarray*}
Now, we need to verify that 
$M = \{\mbox{\sl ap}_2(a_1,a_2,a_3),\mbox{\sl ap}_1(b_1,b_2,b_3)\mid a_3 > b_3\}$ satisfies $T_{P'}(M)\subseteq M$. Using the 2 clauses, this
is reduced to ``$[t_1|t_2] > t_2$'' and
``$t_3 > t_4$ implies $[t_5|t_3] > [t_5|t_4]$'', for any terms 
$t_1, t_2, t_3, t_4$ and $t_5$, 
imposing monotonicity and
subterm properties on $>$. The proof is completed analogously to the 
{\sl permute} example.
$\hfill\Box$\end{example}

It should be noted that in general unfolding can transform a non-terminating
program to a terminating one by replacing infinite branches of the
LD-tree with failing ones~\cite{Bossi:Cocco}. Bossi and Cocco~\cite{Bossi:Cocco} also stated conditions on unfolding that impose termination to be preserved.

\section{Conclusion}
We have presented a non-transformational approach to termination analysis
of logic programs, based on general orderings. The problem of termination
was studied by a number of authors (see~\cite{DeSchreye:Decorte:NeverEndingStory} for the survey). More recent work on this topic can be found among others
in~\cite{Decorte:DeSchreye:98,Decorte:DeSchreye:Vandecasteele,
Etalle:Bossi:Cocco,Hoarau:thesis,Lindenstrauss:Sagiv,Ruggieri:thesis,Smaus:thesis,Taboch,Verbaeten:thesis}. 
The transformational approach to termination has been studied among others
in~\cite{Aguzzi:Modigliani,Arts:PhD,Ganzinger:Waldmann,Krishna:Rao,Marchiori}

Our approach gets its power from integrating the traditional 
notion of acceptability~\cite{Apt:Pedreschi:Studies} with the
wide class of orderings that have been studied in the context of the
term-rewriting systems. In theory, such an integration is unnecessary: acceptability (based on level mappings only) is already equivalent to LD-termination.
In practice, the required level mappings may sometimes be very complex (such
as for Example~\ref{example:rep:der} or Example~\ref{example:dist}~\cite{Dershowitz:Hoot}, {\sl boolean ring}~\cite{Hsiang} or {\sl flattening of a binary tree}~\cite{Arts:PhD}), and automatic  systems for 
proving termination are unable to generate them. In 
such cases, generating an appropriate ordering, replacing the level
mapping, may often be much easier, especially since we can reuse the impressive
machinery on orderings developed for term-rewrite systems. In some other
cases, such as {\sl turn}~\cite{Bossi:Cocco:Fabris}, simple level 
mappings do exist (in the case of {\sl turn}: a norm counting the number of 0s 
before the first occurrence of 1 in the list is sufficient), but most 
systems based on
level mappings will not even find this level mapping, because they only 
consider mappings based on term-size or list-length norms. 
Meta-interpreters, as illustrated in Example~\ref{example:solve}, give 
the same complication.
Again, our approach is able to deal with such cases.

Sometimes level mappings and norms provide an advantage over general 
orderings. This is mostly the case if the termination proof can benefit
from arguments based on arithmetical operations on the numerical values 
provided by the level mapping and norm, as illustrated in Example~\ref{example:conf}. Note however, that general orderings include orderings based on 
mappings and norms as a special case. We can allow the latter types of orderings
as a special case, resorting to them when other orderings in our workbench
fail to produce a proof. If we do resort to them, we may allow arithmetic
operations on them. The main reason why we defined interargument relations in
a very general way is exactly to allow all the power of numerical orderings, and
arithmetic, to be applicable in our context.  

Unlike transformational approaches, that establish the termination results
for logic programs by the reasoning on termination
of term-rewriting systems, we apply the orderings directly to the logic
programs, thus, avoiding transformations. This could both be regarded as an 
advantage and as a drawback of our approach. It may be considered as a 
drawback, because reasoning on successful instances of intermediate body-atoms
introduces an additional complication in our approach, for which there is no
counterpart in transformational methods (except for the transformation step
itself). On the other hand, we consider it as an advantage, because it is
precisely this reasoning on intermediate body atoms that gives more insight 
in the property of {\em logic program termination\/} 
(as opposed to {\em term-rewrite system
termination}). Another advantage over transformational approaches is  
that most of these are restricted to well-moded programs and goals, while
our approach does not have this limitation. 

So, in a sense our approach provides the best of both worlds: a
means to incorporate into `direct' approaches the generality of general
orderings.

We consider as a future work a full implementation of the approach.
Although we already tested very many examples manually, an implementation 
will allow us to conduct a much more extensive experimentation, comparing
the technique also in terms of efficiency with other systems. Since we apply
a demand-driven approach, systematically reducing required conditions to more
simple constraints on the ordering and the model, we expect that the method can
lead to very efficient verification.

\section{Acknowledgements}
We thank Robert Kowalski for continuously stimulating us to look outside of our
ivory tower of research to search for challenges in cross-fertilisation of 
different streams of work.

Alexander Serebrenik is supported by GOA: ``${LP}^{+}$: a second generation
logic programming language''. We thank Maurice Bruynooghe for useful 
suggestions.

\bibliography{/home/alexande/M.Sc.Thesis/main}
\bibliographystyle{abbrv}
\end{document}